\documentclass{cjpsuppl}
\usepackage{graphicx}

\newcommand{\ra}{\mbox{$\rightarrow$}}

\newcommand{\zpr}{\mbox{$Z'$}}

\newcommand{\als}{\mbox{$\alpha_s$}}

\newcommand{\skipblk}[1]{}                                                      
\def\bqa{\begin{eqnarray}}                                                      
\def\eqa{\end{eqnarray}}                                                        

\newcommand{\sto}{\mbox{$SU(2) \x U(1)$}}                                       
                                                                                       
\newcommand{\x}{\mbox{$\times$}}

\newcommand{\sinn}{\mbox{$\sin^2\theta_W\,$}}

\newcommand{\beq}{\[}                                             
\newcommand{\eeq}{\]}

\def\mxth{\mathsurround=0pt }
\def\xversim#1#2{\lower2.pt\vbox{\baselineskip0pt \lineskip-.5pt
  \ialign{$\mxth#1\hfil##\hfil$\crcr#2\crcr\sim\crcr}}}             
                                    
\def\simle{\mathrel{\mathpalette\xversim <}}                                    

\begin{document}
\title{Status and Phenomenology of the Standard Model}
\authori{Paul Langacker}
\addressi{Department of Physics, University of Pennsylvania, Philadelphia, PA 19104, USA}
\authorii{}    \addressii{}
\authoriii{}   \addressiii{}
\authoriv{}    \addressiv{}
\authorv{}     \addressv{}
\authorvi{}    \addressvi{}
\headtitle{Status and Phenomenology of the Standard Model \ldots}
\headauthor{Paul Langacker}
\lastevenhead{Paul Langacker: Status and Phenomenology of the Standard Model \ldots}
\refnum{}
\daterec{15 January 2005}
\suppl{A}  \year{2005} \setcounter{page}{1}
\maketitle

\begin{abstract}
The status of the new standard model is briefly surveyed, with emphasis on
experimental tests, unique features, theoretical problems, necessary
extensions, and possible TeV signatures of Planck scale physics.
\end{abstract}

\section{The {\em New} Standard Model}
The New Standard Model (NSM)\footnote{For a general review, see~\cite{pdg}.} is the original 
SM ($SU(3) \x SU(2) \x U(1) \x$ {  classical general relativity), 
supplemented with neutrino mass, which can be Dirac or Majorana.
The focus of this talk will be on the electroweak (\sto) sector.

The NSM is a mathematically consistent field theory of the strong, weak, and electromagnetic
interactions and classical gravity that is the correct description of nature
to first approximation down to $\sim10 ^{-16}$ cm. However, it is a
complicated theory with many free parameters and fine tunings, strongly
suggesting that there must be new physics  beyond the standard model (BSM)
at shorter distance scales, and in fact most
particle physicists are involved in one way or another in the search for that new physics.

The standard model has many special features that are {\em  usually not} maintained in BSM,
leading both to a challenge for theorists in constructing new models and an opportunity
for experimentalists searching for hints of new physics.
These include:
\begin{itemize}
\item The neutrino masses $m_\nu$ were predicted to vanish in the {\em old} standard model.
 One would need to add singlet fermions and/or triplet Higgs fields and/or higher dimensional
 operators (HDO) for nonzero masses.  However, almost all extensions of the SM
 predicted one or more of these.
\item Yukawa couplings  are proportional to fermion masses,
 $h_f \propto g m_f/M_W$, implying that the Higgs couplings are small for the light
fermions and that they are flavor conserving.  
This is partially maintained in the Minimal Supersymmetric Standard Model (MSSM) and simple 
two Higgs doublet models but usually not in more complicated Higgs models. 
\item There are no flavor changing neutral currents
(FCNC) mediated by the $Z$, $\gamma$, or Higgs ($H$) at tree level.
Lepton flavor is conserved to all orders for massless neutrinos (and the {\em direct}
effects of $m_\nu \ne 0$ for charged lepton processes are negligible).
Quark FCNC at loop level are  suppressed (the GIM mechanism).
There are enhanced FCNC effects in most extensions of the SM, including
new loop effects in supersymmetry, new interactions in dynamical symmetry breaking,
Little Higgs models, 
possible heavy $Z'$ bosons with non-universal couplings (motivated in some string
constructions), and multiple Higgs doublets.
\item Off-diagonal $CP$ violation (e.g., in $K$ and $B$ mixing and decays) is
suppressed, while diagonal $CP$ violation, i.e., electric dipole moments
 (EDM), is highly suppressed.
Extensions again typically offer new sources of enhanced $CP$ violation, 
such as possible new $\not \! \! CP$ phases in supersymmetry ($\mu$ and soft 
parameters, and the associated supersymmetric loops), 
and phases associated with additional Higgs and 
exotic fields or new non-universal $Z'$ bosons.
\item Baryon ($B$) and lepton ($L$) number are conserved perturbatively 
(and $B-L$ non-perturbatively) in the SM, while they can be violated in BSM,
such as grand unified theory (GUT) or string interactions, possible $R$-parity 
violation ($\not \! R_p$) in supersymmetry, and Majorana neutrino masses. 
\end{itemize} 



\section{Experimental Precision Tests}
Let me briefly comment on some of the experimental tests
of the standard electroweak model.

\subsection{The Weak Charged Current}
Like QED, the Fermi Theory of the weak charged current interactions 
had been developed and tested prior to the standard model. However, 
it was successfully
incorporated and  greatly improved by being made renormalizable,
which allowed the calculation of radiative corrections. Much of the recent 
activity has involved study of the Cabibbo-Kobayashi-Maskawa
(CKM) matrix 
\beq V = \left( \begin{array}{ccc} V_{ud} & V_{us} & V_{ub} \\
V_{cd} & V_{cs} & V_{cb} \\ V_{td} & V_{td} & V_{td} \end{array}
\right), \eeq 
which measures the mismatch between the
family structure of the left-handed $u$-type and $d$-type quarks.
 For 3 families $V$ involves three angles and one $CP$-violating
phase after removing unobservable $q_L$ phases. The corresponding
right-handed quark mixings are unobservable in the SM, but could be
if there were new interations involving $q_R$. There have been 
extensive recent studies, especially in $B$ and $K$ decays~\cite{giorgi,sakai,patera}, to test the
unitarity and consistency of $V$ as a probe of new physics and to test the origin of
CP violation\footnote{An
additional source of CP breaking is required for baryogenesis.}.
The longstanding hint of a 2.4$\sigma$ discrepancy in the universality
prediction $|V_{ud}^2|+|V_{us}^2|+|V_{ub}^2| = 1$ has recently been resolved
by new measurements of $|V_{us}|$ by the BNL-E865, Fermilab KTeV, CERN NA48,
and $DA\Phi NE$ KLOE experiments~\cite{patera}, which yield a higher
value than the previous world average.

\begin{figure}[htbp]
\begin{center}
\includegraphics*[scale=0.6]{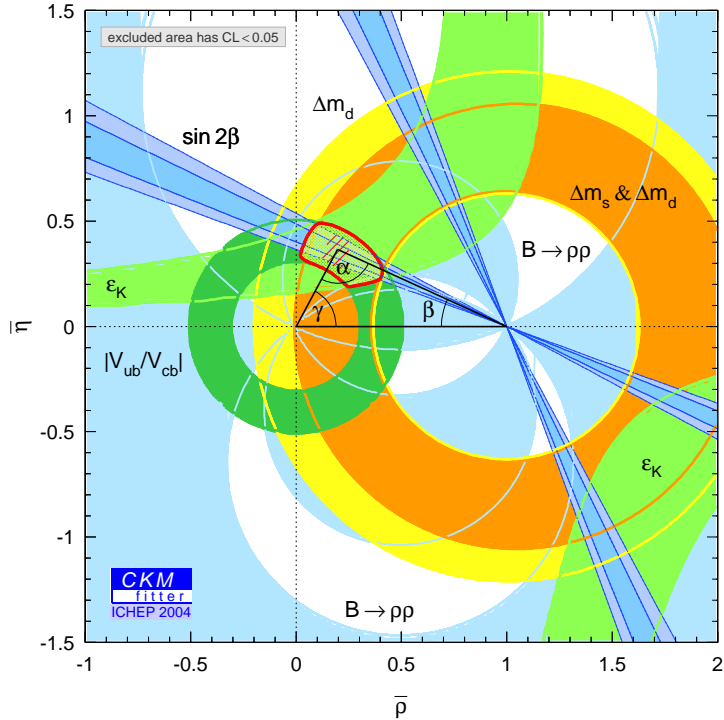}
\caption{The current status of the unitarity triangle, from~\cite{ckmfitter}.}
\label{ckmfitter}
\end{center}
\end{figure}
The sides of the unitarity triangle are well
determined~\cite{ckmfitter} by $B$ decay rates, branching ratios, and oscillations (an actual measurement of
$B_s$ oscillations in eagerly awaited), and by $CP$ violation
in $K$ decays. The focus has turned to a measurement of the angles
from $CP$ violation in $B$ decays in order to overconstrain the system. The
Babar and Belle collaborations~\cite{giorgi,sakai} have
obtained the very precise value $\sin 2 \beta = 0.726 \pm 0.077$ from $B_d^0(t) \ra J/\psi K_{S,L}$,
 with little theory error. 
The angles  $\alpha$ and $\gamma$ are more difficult. Recently, direct $CP$
violation in charmless $B$ decays has been established, and precise
values for $\alpha$ (e.g., an average $106^{\circ +8^\circ}_{-11^\circ}$~\cite{giorgi})
are emerging. Babar and Belle are now in good agreement on their measurements of
$b\rightarrow s$ electroweak penguin decays, such as $B^0 \ra \phi K^0$. There is
some evidence (at about the 2.5$\sigma$ level) for a discrepancy with the SM expectation.
If this were confirmed by future higher precision measurements, it would
suggest such new physics as supersymmetric loops 
(for high $\tan \beta$)~\cite{susyloops}
or a heavy $Z'$ gauge boson with flavor changing couplings~\cite{zprimefcnc}. 
The latter would be
a tree level effect competing with a small SM loop diagram.

\begin{figure}[htbp]
\begin{center}
\includegraphics*[scale=0.4]{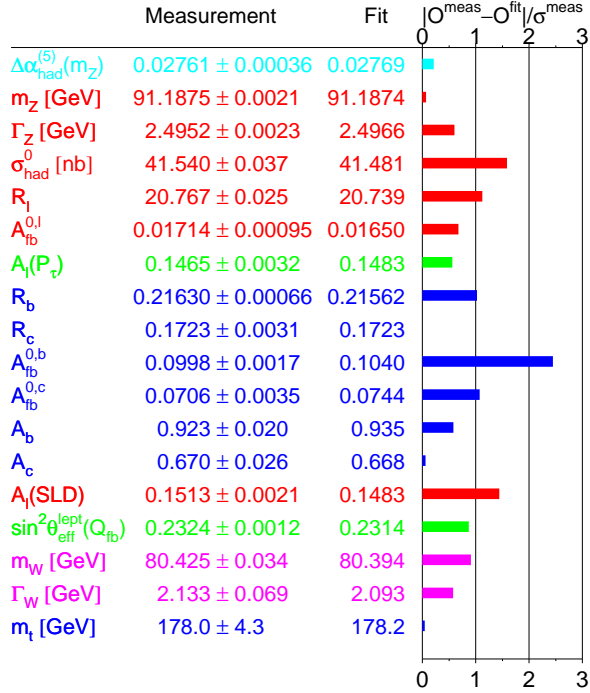}
\caption{Precision observables, compared with their expectations from the
best SM fit, from~\cite{lepewwg}.}
\label{pulls}
\end{center}
\end{figure}
\subsection{The Weak Neutral Current}
The weak neutral current (WNC), along with the
$W$ and $Z$, have been the   primary predictions and tests  of the electroweak 
\sto \ model.
The WNC was  discovered in 1973 by the Gargamelle 
collaboration at CERN and by HPW at Fermilab.
The structure of the WNC has been tested  in many processes~\cite{wnc}, including
(purely weak) neutrino scattering
 $ \nu e \ra
\nu e, \; \nu N \ra \nu N, \; \nu N \ra \nu X$; weak-electromagnetic
interference in $e^{\uparrow \downarrow} D
\ra e X$,  atomic parity violation, and recently in polarized M\"oller scattering~\cite{moller}; 
and in $e^+ e^-$ scattering
above and below the $Z$ pole.

There is generally excellent agreement of the WNC data with the SM predictions.
An apparent discrepancy in the precise  Boulder measurement of parity violation
in the cesium atom was resolved by improved calculations, especially of the radiative corrections~\cite{apvreview}.
There is still an $\sim 3 \sigma$ discrepancy between the NuTeV ratio of neutral to
charged current deep inelastic $\nu N$ cross sections and the SM expectation~\cite{nutev}.
This could possibly be due to new TeV scale effects such as a heavy $Z'$, but may well be
a subtle strong interaction effect, such as an $s-\bar s$ asymmetry or large isospin
breaking in the nucleon sea, or higher order QCD effects~\cite{nutevtheory}.

\subsection{The LEP/SLC Era}
The $Z$ factories LEP and SLC allowed tests of the standard model
at a precision of $\sim 10 ^{-3}$, much greater than had previously been possible
at high energies. In particular, the four LEP experiments ALEPH,
   DELPHI, L3,  and OPAL at CERN accumulated some $2 \times 10^{7} Z's$
   at the $Z$-pole in the reactions  $e^+ e^- \rightarrow Z \rightarrow \ell^+ \ell^-$ and
  $q \bar{q}$. The SLD experiment at SLAC had  a smaller number of events, 
  $\sim 5 \times 10^5 $, but had the significant advantage of the highly polarized
 ($\sim$ 75\%)  SLC $e^-$ beam.
The $Z$ pole observables included the lineshape variables,
 $M_Z, \Gamma_Z, $ and  $\sigma$; and the branching ratios into $e^+e^-, \mu^+ \mu^-, \tau^+ \tau^-$
 as well as into  $q \bar{q}, c \bar{c}, b \bar{b},$ and (less precisely) $ s \bar{s}$.
 These could be combined to obtain the stringent constraint
$N_\nu = 2.9841 \pm 0.0083$ 
on the number of ordinary neutrinos with
$m_\nu < M_Z/2$ (i.e., on the number of families with a light $\nu$).
This also constrains other invisible $Z$ decays, such as into a light scalar $\nu$ in
supersymmetry. The $Z$-pole experiments also measured a number of asymmetries,
including forward-backward (FB), polarization, the $\tau$ polarization, and mixed 
FB-polarization, which were especially useful in determining the weak angle \sinn.
The leptonic branching ratios and asymmetries confirmed the lepton family universality
predicted by the SM. The results of many of these observations, as well
as some WNC and high energy collider data, are shown in Figure \ref{pulls}.
There is generally excellent agreement with the SM expectations, though there is a
hint of a tension between the lepton and quark asymmetries (most apparent in
the $b$ quark forward-backward asymmetry $A_{fb}^{0,b}$ and the polarization
asymmetry $A_l$.). This may well be a fluctuation, but 
could possibly be suggesting new physics affecting the third family.
 
The LEP II program above the $Z$-pole provided a precise determination of $M_W$ (as
did the Tevatron experiments CDF and D0), measured the four-fermion cross
sections $e^+e^- \ra f \bar f$, searched for the Higgs and for BSM effects
such as superpartners, and tested the  (gauge invariance) predictions 
of the SM for the three and four point gauge self-interactions.
These had already been indirectly verified by the  \als \ running in QCD
and by the electroweak radiative corrections, but could be probed more directly
in  $e^+e^- \ra W^+W^-$. The SM predicts strong cancellations between
the three tree-level diagrams in Figure \ref{eeww}  in the
high energy amplitude, which  would be upset
by anomalous couplings. As seen in Figure \ref{sigww} the LEP II
results are in excellent agreement with the SM prediction.
\begin{figure}
\begin{center}
\includegraphics*[scale=0.5]{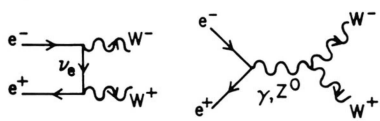}
\caption{Tree-level diagrams contributing to $e^+e^- \ra W^+W^-$}
\label{eeww}
\end{center}
\end{figure}

\begin{figure}[htbp]
\begin{center}
\includegraphics*[scale=0.5]{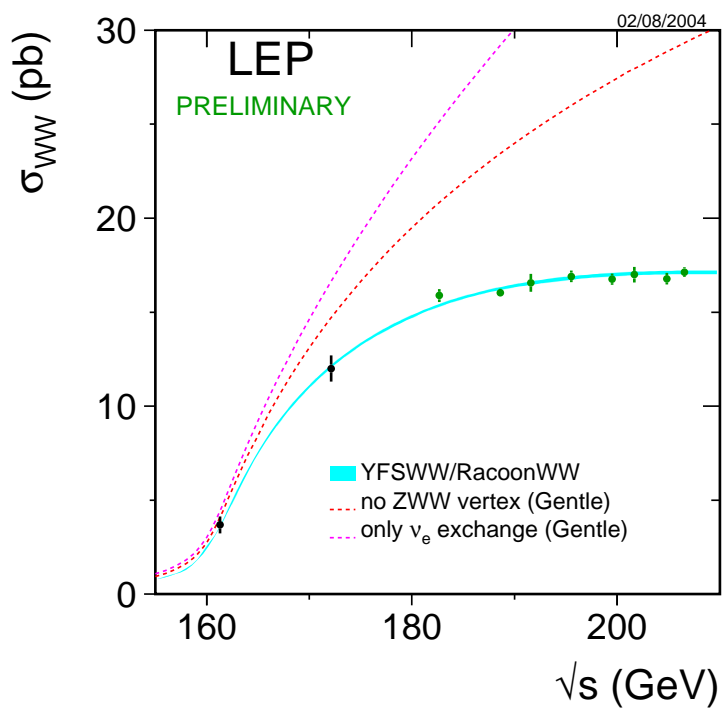}
\caption{The $e^+e^-\ra W^+W^-$ cross section, compared with the SM
expectation, from~\cite{lepewwg}.}
\label{sigww}
\end{center}
\end{figure}

\subsection{The Anomalous Magnetic Moment of the Muon}
The  BNL E821 experiment has made an extremely precise measurement of
the muon anomalous magnetic moment~\cite{bnl821},
$ a_\mu(\mbox{exp}) = (11659208.0 \pm 5.8) \times 10^{-10}$, which is
$2.4 \sigma$ above the SM prediction~\cite{gm2theory} assuming the value for the hadronic vacuum polarization  obtained from
$e^+e^- \ra$  hadrons: one finds  $23.9 (7.2) (3.5) (6) \times 10^{-10}$ for
$ \Delta a_\mu \equiv a_\mu(\mbox{exp}) - a_\mu(\mbox{SM})$.
The discrepancy is smaller ($\sim 0.9\sigma$) using the value
calculated from $\tau$ decays. If the discrepancy is real, it could be accounted for
in the MSSM by loops involving a scalar muon and a neutralino, or a scalar neutrino
and a chargino. One has~\cite{gm2theory} 
$ \Delta a_\mu (\mbox{SUSY}) \sim 13 \times 10^{-10} \frac{\tan\beta~ \mbox{sign}(\mu)}{(M_{\mbox{\tiny SUSY}} / 100 \mbox{\small ~GeV})^2},$
for common superpartner masses $M_{\mbox{\tiny SUSY}}$, favoring positive $\mu$
(the supersymmetric Higgsino mass),
large $\tan \beta$  
 and/or small $M_{\mbox{\tiny SUSY}}$.

\subsection{The Precision Program}
Altogether, the precision program, including the WNC, the $Z$ and $W$, 
the $Z$-pole and above, and the Tevatron measurements of  $m_t$,
have shown that
  \begin{itemize}
   \item The SM is correct and unique to zeroth approximation, justifying the gauge
    principle as well as the SM gauge group and fermion representations. 
   \item The SM is correct at the loop level, confirming the framework of renormalizable 
    gauge field  theory, leading to successful predictions for $m_t$ and $\alpha_s$,
    and to a still untested prediction for the Higgs mass
    $M_H$. The indirect (precision) data leads to the prediction $M_H = 113^{+56}_{-40}$ GeV,
    consistent with the direct lower limit $M_H > 114.4$ GeV
     from the nonobservation of $e^+e^-\ra ZH$ at LEP II\footnote{ 
    The central value is pulled up by $A_{fb}^{0,b}$  and down by $A_l$. 
    It has increased somewhat due to the
    new D0 Run I $m_t$ value~\cite{d0mt} and a lower $M_W$.},
    while the indirect and direct data together  imply $M_H <$246 GeV at 95\% cl~\cite{wnc,lepewwg}.
    This is consistent with, but does not prove, the MSSM prediction of a light Higgs, e.g.,
    $M_H \simle 130$ GeV~\cite{susyloops}.
   \item Possible new TeV-scale physics is severely constrained.
   In particular BSM physics which decouples, such as supersymmetry and
   unification, is strongly favored over non-decoupling physics, such as most
   forms of compositeness.
   \item The precisely measured gauge couplings are consistent with the
   gauge unification~\cite{GUT} predicted by the MSSM. If this is not an accident or
   due to a compensation, it strongly limits the possibilities for new TeV scale physics.
  \end{itemize}

\section{Problems With the New Standard Model}
Despite its successes, the new standard model has a great  deal of
arbitrariness and fine-tuning~\cite{ssm},  as is illustrated by the fact that
it has 27 free parameters (29 for Majorana neutrinos), {\em not}
including electric charges. These can be taken to be
 3 gauge couplings; the $Z$ and Higgs masses; the QCD
 $\theta$ parameter;  12 fermion masses; 6 mixings and 2 CP phases (2 additional
 for Majorana $\nu$'s); and  the cosmological constant. The Planck scale (Newton
 constant) is not included because only the ratios of mass parameters are observable.
 In particular,
 \begin{itemize}
\item The SM gauge group is   complicated: it involves 3 distinct gauge couplings,
only the electroweak part is parity-violating,
and charge quantization (e.g., $|q_e| = |q_p|$) is put in by hand
(anomaly cancellation by itself is not sufficient to determine all of the
hypercharge assignments). This suggests some form of unification,
such as grand unification (GUT) or string theory.

\item The existence of fermion families, the hierarchies of their masses,
and the pattern of their mixings are unexplained in the NSM. 
It is still unknown whether the neutrinos are Dirac or Majorana, and 
their  large mixings and small mass scale compared to other fermions
are not understood. The fermion spectrum may well be
probing the Planck or GUT scale. In string theories the hierarchies
may be associated with higher dimensional operators in heterotic models
or with volume-suppressed
instanton effects in intersecting brane constructions.
Other possibilities include compositeness,
family symmetries, radiative hierarchies, or large extra dimensions.
Also, the CP violation in the NSM is inadequate to explain the observed
baryon asymmetry, suggesting new sources of CP violation, such as
new CP phases associated with soft breaking or $\mu$ parameters in supersymmetric
models of electroweak baryogenesis,
or new neutrino phases in leptogenesis.

\item
Consistency of the SM requires a Higgs mass-squared
comparable to the electroweak scale, 
$M_H^2 = O(M_Z^2)$. However, 
quadratically divergent  higher order corrections, such as shown in Figure \ref{hot},
 yield much  larger corrections, $\delta M_H^2/M_Z^2 \sim 10^{34}$,
 where I have assumed the integrals are cut off at the Planck scale,
 requiring an extremely fine-tuned cancellation with the tree-level mass-squared.
 \begin{figure}[htbp]
\begin{center}
\includegraphics*[scale=0.5]{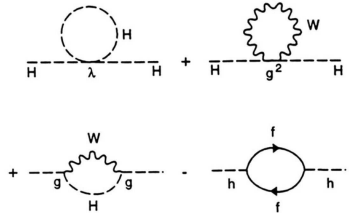}
\caption{Quadratically divergent corrections to $M_H^2$.}
\label{hot}
\end{center}
\end{figure}
The traditional solutions to this fine-tuning problem include TeV-scale
supersymmetry~\cite{susyloops}, in which fermion and boson loops cancel; Little Higgs
models~\cite{lhiggs}, in which fermions and bosons cancel separately;
dynamical supersymmetry breaking (DSB)~\cite{dsb}, in which there are
no elementary Higgs fields; and large extra dimensions~\cite{led}, in which the
fundamental scale is much lower. Recently, consideration has been given to
theories, such as split supersymmetry~\cite{split}, in which supersymmetry is
broken at a high scale and the Higgs hierarchy problem is apparently solved
by a fine tuning, perhaps related to an enormously large ``landscape'' of
superstring vacua~\cite{landscape} and possible anthropic~\cite{anthropic}
 selection principles.

  \item The strong CP problem~\cite{strongcp} refers to the fact that 
  one can add the P, T, and CP-violating term
 $\frac{\theta}{32 \pi^2} g_s^2 F \tilde{F}$ to the QCD Lagrangian,
 where $\tilde{F}_{\mu \nu} = \epsilon_{\mu \nu \alpha \beta} F^{\alpha
\beta}/2$ is the dual field and $\theta$ is an arbitrary dimensionless parameter.
The experimental bound on the neutron electric dipole moment implies
 $\theta < 10^{-9}$. One cannot simply set $\theta$ to zero because weak
 interaction corrections shift $\theta$ by
  $\delta \theta|_{\rm  weak} \sim 10^{-3}$,
  again requiring a fine-tuned cancellation between the tree and weak contributions,
  for which (as far as I am aware)
  there is no anthropic explanation.
 Possible solutions include the extension of the SM to include a
  spontaneously broken global $U(1)$ (Peccei-Quinn) symmetry.
 This implies the existence of an axion, for which there is still
 a narrow window for the breaking scale allowed by cosmological and astrophysical
 considerations~\cite{darkmatter} (which is small compared to the 
 scale expected in simple string implementations~\cite{straxion}).
   Other possibilities include an unbroken global  $U(1)$, leading to a
   (disfavored) massless $u$ quark; or 
   spontaneously broken CP (leading to possible cosmological domain
   wall problems) combined with other symmetries so that $\theta$ becomes calculable and 
   small.  

  \item Gravity is not unified with the other interactions. Furthermore, although classical general
  relativity is included in the SM, it is
 not renormalizable when quantized. Finally, the vacuum energy
 $ \langle V \rangle$ from electroweak symmetry breaking leads
 to an effective
  cosmological constant: $\Lambda_{\rm SSB} = 8 \pi G_N
   \langle V \rangle$ some $ 10^{50} $ times larger than the
   observed value from the acceleration of the universe,
   requiring an extremely finely-tuned cancellation with the primodial
   value. The tuning is even worse ($\Lambda_{\rm SSB} > 10^{124} 
   \Lambda_{\rm  obs}$) in simple GUT and
   string constructions. This difficulty is made even worse in that
   it is not enough to somehow eliminate the string scale contribution - one must 
   simultaneously eliminate the electroweak and the smaller QCD contributions, as
   well as loop effects.
The  unification problem is solved in supergravity and Kaluza Klein
theories. The extension to superstrings renders quantum gravity renormalizable
(in fact finite). There is no known solution to the cosmological constant problem
other than an anthropically motivated fine-tuning associated with the
string landscape~\cite{landscape}.

\end{itemize}

\subsection{Necessary New Ingredients}
New observations as well as the SM problems imply the need for
a number of new ingredients. These include
  \begin{itemize}
\item A mechanism for small neutrino masses~\cite{neutrino}. 
More generally, one is interested in whether the neutrino masses are
Dirac or Majorana, whether the spectrum is ordinary or inverted, what
the absolute scale is (with implications for cosmology), why two mixing angles
are large, the value of the small angle $U_{e3}$ and the CP violating phase,
whether the LSND results (which suggest the existence of additional, sterile, neutrinos
which mix with the ordinary neutrinos) are confirmed by MiniBooNE,
and whether there are any new $\nu$ interactions or anomalous properties.
Mechanisms for neutrino mass include some form of
seesaw mechanism\footnote{The canonical seesaw may not be
favored in some superstring constructions~\cite{giedt} or models with a TeV scale
$Z'$~\cite{zprimenu}.}, perhaps associated with the Planck or GUT scale; heavy Higgs triplets; 
or small Dirac masses, e.g., from higher-dimensional operators or volume
suppressions in theories with LED.
\begin{figure}[htbp]
\begin{center}
\includegraphics*[scale=0.5]{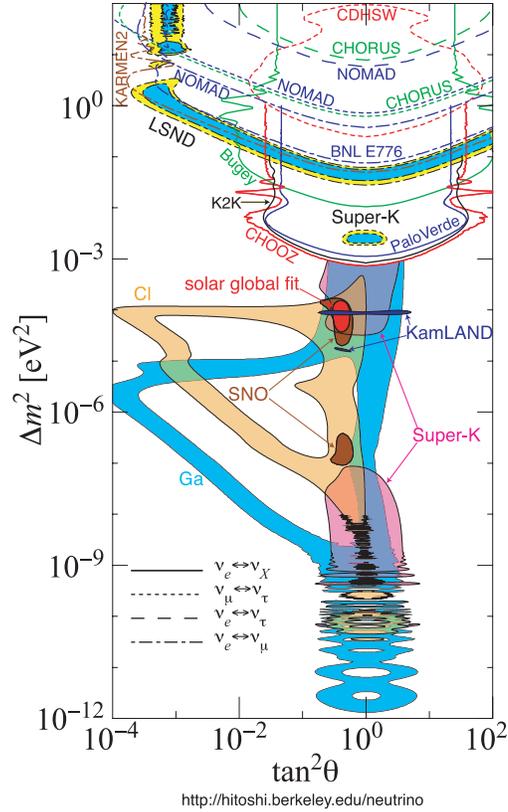}
\caption{Current neutrino oscillation results, from~\cite{hm}.}
\label{nuosc}
\end{center}
\end{figure}
\item A mechanism for generating the baryon asymmetry~\cite{baryogenesis}, such as
electroweak baryogenesis, probably (except for a small parameter range~\cite{mssmbary})
requiring an enhancement with respect to the
MSSM from an extended Higgs sector~\cite{higgsbary} and/or a \zpr~\cite{zprbary}; 
heavy Majorana neutrino decays (leptogenesis), associated with canonical seesaw
models; the decay of a coherent field; or  CPT violation.
\item A mechanism for the dark energy~\cite{darkenergy}, which constitutes about
70\% of the energy density of the Universe. Is it a cosmological constant
(and if so, why is it so small),
or some form of time-varying field (quintessence)? This raises other questions,
such as whether it is related to inflation, and whether there is 
a connection to the (controversial) suggestion of time variation of 
couplings~\cite{time}?
\item An explanation for the nature of the dark matter~\cite{darkmatter} (30\%),
such as the lightest supersymmetric particle (if $R$-parity is conserved) or an axion.
\begin{figure}[htbp]
\begin{center}
\includegraphics*[scale=0.45]{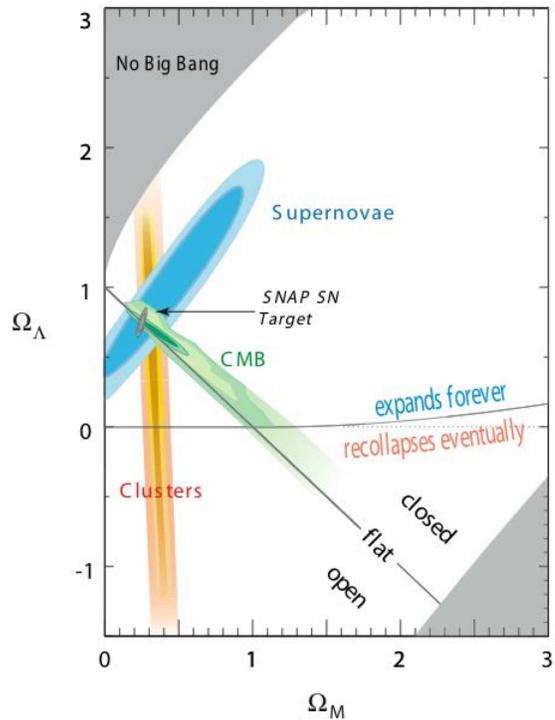}
\caption{Current constraints on the dark energy ($\Omega_\Lambda$) and total
matter density ($\Omega_M$), most of which is dark, from~\cite{omlam}.}
\label{omlam}
\end{center}
\end{figure}
 \item Mechanisms to suppress  flavor changing neutral currents~\cite{fcnc,fcnc2}, 
 proton decay~\cite{GUT,proton}, and
 electric dipole moments~\cite{fcnc2,edm}. These are automatically absent or strongly suppressed in
 the SM, but are present at some level in almost all extensions.
 There are experimental opportunities to greatly improve limits or observe
  rare $\mu,\ \tau,\ B$, and $K$ decays (or $\mu-e$ conversion); neutron, electron, or atomic EDMs;
  and proton decay. These are extremely important complements to the future collider program.
  \end{itemize}

\subsection{TeV Physics from the Top Down}
Speculations concerning new physics at the TeV scale are often motivated
by attempts to solve the problems of the standard model. For example, supersymmetry,
dynamical symmetry breaking, Little Higgs models, and large extra dimensions (LED) are all
motivated at least in part by the fine tuning associated with the Higgs mass in the SM.
It is entirely possible, however, that there may be new TeV scale physics in addition to
(or instead of) the above that does not directly solve any SM problem, but which simply
emerges as a remnant of the underlying physics at a much higher mass scale~\cite{topdown}.
For example, concrete semi-realistic superstring constructions often lead to such
new physics as: (a) Additional $Z'$ gauge bosons or other new interactions. These are especially
common~\cite{zprimerev}, not only in string constructions but also in dynamical
symmetry breaking~\cite{dsb}, Little Higgs~\cite{lhiggs}, and LED~\cite{led}.
New $Z'$s  may have implications for a highly non-standard Higgs sector, baryogenesis,
cold dark matter, and FCNC. (b) New exotic particles are very common.
These may include additional Higgs singlets and Higgs doublet pairs (all of which can
mix, and which may yield new sources of CP violation and FCNC), quarks and leptons
with non-canonical weak interactions (e.g., heavy charge $-1/3$ quarks $D_{L,R}$, both
of which are $SU(2)$ singlets~\cite{exotic}), or even fractional electric charges
such as $\pm 1/2$
(the latter usually couple also to a hidden sector and {\em may} be confined).
(c) Quasi-hidden sector gauge groups. If strongly coupled these may be associated
with supersymmetry breaking by gaugino condensation. However, they are not
always strongly coupled. There are usually a few particles that couple to both the ordinary
and hidden sectors (hence the term quasi). Extra $U(1)'s$ also often couple to both sectors.

Of course, such things may simply be flaws in the constructions studied, but they may also
be hinting that there really is TeV physics beyond the MSSM. In particular, superstring
model builders should not necessarily assume that the ultimate goal is to obtain the MSSM.

\section{Conclusions}
The Standard Model is spectacularly successful, 
and successfully describes nature down to a distance scale a thousandth the size
of the atomic nucleus. However, the SM has many parameters, tunings, and
unexplained features, indicating that there must be underlying 
new physics that manifests itself on
shorter distance scales. There are many theoretical ideas,
including superstrings,
Grand Unification~\cite{GUT} (including canonical GUTs in 4 dimensions and modified 
versions which only
fully manifest themselves in higher dimensions, e.g., within a string theory),
and supersymmetry. Even Planck or GUT-scale physics may leave telltale
signatures at the TeV scale, such as $Z'$s, extended Higgs sectors, and exotic particles.
Other theoretical ideas include large extra dimensions (and deconstruction),
dynamical symmetry breaking,
compositeness, and Little Higgs models.

The TeV scale may well be extremely complicated. Fortunately, we can look
forward to a variety of  experimental probes. These include
hadron colliders (the Tevatron and LHC) and linear colliders (the International Linear
Collider and CLIC) to explore the energy frontier. In addition, there will be 
a (very important) complementary program of lower energy probes, including
detailed explorations of off-diagonal CP violation, rare decays, and FCNC
at $B$ factories and in $K$, $\mu$, and $\tau$ decays; more precise measurements
of $n$, $e$, and atomic EDM; precision experiments; 
 neutrino physics; $p$ decay; and new observations in
cosmology/astrophysics, including ground and space based studies of dark matter and
dark energy.

 There are tremendous opportunities in particle physics. There are theoretical ideas
 and experimental facilities that could possibly - if we are extremely lucky -
 allow us to develop and test a new standard theory
valid all the way to the Planck scale.

\bigskip

{\small This work was supported by the U.S. Department of Energy grant 
DOE-EY-76-02-3071.}


\end{document}